\documentclass[12pt]{iopart}
\usepackage{hyperref}
\usepackage{tikz}
\usepackage{iopams}
\usepackage{graphicx}
\newcommand*\diff{\mathop{}\!\mathrm{d}}
\usepackage{bm}
\usetikzlibrary{arrows.meta}
\usepackage{braket}
\usepackage[caption=false]{subfig}
\begin{document}

\title[Optimal control of mirror pulses for cold-atom interferometry]{Optimal control of mirror pulses for cold-atom interferometry}
\author{Jack Saywell$^1$, Ilya Kuprov$^2$, David Goodwin$^2$, Max Carey$^1$ and Tim Freegarde$^1$}
\address{$^1$ School of Physics \& Astronomy, University of Southampton, Highfield, Southampton, SO17 1BJ, UK}
\address{$^2$ School of Chemistry, University of Southampton, Highfield, Southampton, SO17 1BJ, UK}
\ead{j.c.saywell@soton.ac.uk}
\date{\today}

\begin{abstract}Atom matterwave interferometry requires mirror and beamsplitter pulses that are robust to inhomogeneities in field intensity, magnetic environment, atom velocity and Zeeman sub-state. Pulse shapes determined using quantum control methods offer
significantly improved interferometer performance by allowing broader atom distributions, larger interferometer areas and higher contrast.
We have applied gradient ascent pulse engineering (GRAPE) to optimize the design of phase-modulated mirror pulses for a Mach-Zehnder light-pulse atom interferometer, with the aim of increasing fringe contrast when averaged over atoms with an experimentally relevant range of velocities, beam intensities, and Zeeman states. Pulses were found to be highly robust to variations in detuning and coupling strength, and offer a clear improvement in robustness over the best established composite pulses. The peak mirror fidelity in a cloud of $\sim 80\ \mu$K ${}^{85}$Rb atoms is predicted to be improved by a factor of 2 compared with standard rectangular $\pi$ pulses.
\end{abstract}

\maketitle

\section{Introduction}

Emerging quantum technologies require the coherent manipulation of quantum states. For example, ultra-precise cold-atom-based sensors such as gravimeters, accelerometers, magnetometers and gyroscopes \cite{Kasevich1991, Gustavson1997, DeAngelis2009, Barrett2014} use interactions with laser pulses to form the beamsplitters and mirrors of matterwave interferometers \cite{RBerman1997}, and these `$\pi/2$' and `$\pi$' pulses must operate with high fidelity if the best sensitivity is to be achieved by using pulse sequences to increase the interferometer area \cite{McGuirk2000} or maximize the entanglement \cite{Islam2008}.

The fidelity of quantum state manipulation deteriorates when there are inhomogeneities in the interaction field, magnetic environment, atomic velocities and quantum state distributions \cite{Dunning2014}. This limits the number of control operations that can be performed before coherence is lost, so it is common to filter the atomic sample to restrict the variations experienced \cite{Kasevich1991a, McGuirk2000, McGuirk2002} by fewer atoms. Inhomogeneities thus limit the interferometer area and sample size, and hence the measurement contrast and sensitivity \cite{Szigeti2012a, Butts2013}.

Various techniques have been developed in the field of NMR spectroscopy to produce control pulses that are robust to variations in the interaction strength and detuning, and such techniques should be applicable to other systems including the effective 2-level schemes of atom interferometry. Shaped pulses \cite{Freeman1998, Luo2016, Fang2018}, rapid adiabatic pulses \cite{Baum1985,Kovachy2012,Bateman2007} and composite pulses \cite{Levitt1981, Levitt1983, Cummins2000, Berg2015} all use complex time-dependent interactions to reproduce the desired operation of a single pulse while compensating for the effects of inhomogeneities. For atom interferometry, McGuirk \textit{et al.} \cite{McGuirk2002} suggested that composite pulses could improve the augmentation pulses within Large Momentum Transfer (LMT) arrangements, and Butts \textit{et al.} \cite{Butts2013} demonstrated that the WALTZ \cite{Shaka1983} composite inversion pulse doubled the sensitivity of a cold Cs atom interferometer.

In a feasibility study of the applicability of composite pulses to cold-atom interferometers, Dunning \textit{et al.} \cite{Dunning2014} analyzed the performance of various established NMR pulse sequences for inversion or ‘mirror’ operations in a thermal cloud of ${}^{85}$Rb in a velocity-sensitive Raman arrangement subject to intensity variations and a distribution over Zeeman sublevels. Although cold-atom arrangements differ from NMR applications in the magnitudes of different inhomogeneities and the correlations between them, most of the schemes tested improved the robustness of the inversion operation to both detuning and coupling strength variations, and there was excellent agreement between the observed performance and that predicted computationally, showing there to be no further significant perturbations or decoherence.

In this paper, we address the computational design and optimization of mirror pulses specifically for atom interferometry. Using the well-known optimal control algorithm GRAPE \cite{Khaneja2005} and the advanced spin dynamics simulation toolbox \textit{Spinach} \cite{Hogben2011}, we have derived pulse shapes that we predict will improve interferometric contrast following a Mach-Zehnder sequence within a $100\ \mu $K cloud of ${}^{85}$Rb atoms by a factor of 1.76 compared with standard rectangular pulses of constant phase and intensity. Our pulse shapes compensate off-resonance and pulse-length errors better than established composite pulse sequences, doubling the peak mirror fidelity when simulated in the $\sigma^+ - \sigma^+$ Raman polarization arrangement for a $\sim 80 \mu $K cloud. It is expected that such optimal pulses should allow for greater interferometric areas, higher contrast, warmer samples, and therefore increased interferometric sensitivities.

\section{Coherent Pulse Theory}

\subsection{Two-level Representation}

The Hamiltonian of a two-level atom interacting with a laser may be written in the rotating frame as \cite{RBerman1997}:
\begin{eqnarray} \label{eq: ham}
\hat{H}_R =  \frac{\hbar}{2} \left( \begin{array}{c c}
\delta & \Omega_R e^{-i\phi_L} \\
\Omega_R e^{i\phi_L}	   &  -\delta
\end{array}
\right).
\end{eqnarray}
Here, $\Omega_R$ is the Rabi frequency, $\phi_L$ is the laser phase, and $\delta(t)$ is the detuning from resonance. The detuning can be expressed as \cite{RBerman1997}:
\begin{eqnarray}
\delta(t) =& - \delta^{AC} + \omega_{L}(t) - \bigg( \omega_{12} + \frac{ \mathbf{p} \cdot {\mathbf{k}}_L}{m} +   \frac{ \hbar | {\mathbf{k}}_L |^2 }{2m}    \bigg).
\end{eqnarray}
$\delta^{AC}$ represents the AC Stark shift, $\omega_{L}(t)$ is the laser frequency, and $\omega_{12}$ is the separation of the two levels. ${\mathbf{k}}_L$ is the laser wavevector, $\hbar | {\mathbf{k}}_{L} |^2 /2m$ is the recoil shift, and ${ \mathbf{p} \cdot {\mathbf{k}}_{L}}/{m}$ is the Doppler shift. The two-level system hence consists of the states $\ket{1, \mathbf{p} }$ and $\ket{2, \mathbf{p}+\hbar {\mathbf{k}}_{L} }$, where 1 and 2 refer to the two atomic states and $\hbar {\mathbf{k}}_{L} $ is the photon impulse. If the beams are kept resonant at the centre of the velocity distribution, e.g. by chirping the frequency to account for acceleration under gravity \cite{Kasevich1991, Luo2016}, the detuning can be written as $|{\mathbf{k}}_{L}|v$, where $v$ is the relative speed of a given atom along the direction of the wavevector. Therefore, the detuning of a given atom will remain approximately fixed throughout the interferometer sequence, and the range of detunings will be due to the momentum distribution and hence temperature of the cloud.

In the Bloch sphere representation, the state vector of any two-level system may be written as \cite{Shore2011}
\begin{eqnarray}
\ket{\psi} = \cos\bigg(\frac{\vartheta}{2}\bigg)\ket{1} + e^{i\varphi}\sin\bigg(\frac{\vartheta}{2}\bigg)\ket{2}
\end{eqnarray}
where $\vartheta$ and $\varphi$ are the polar and azimuthal coordinates of a point on the surface of a unit sphere. Free and driven evolution is expressed in terms of rotations about axes through the centre of the Bloch sphere.
The propagator, 
\begin{eqnarray}
\hat{U}(t_1,t_2) = {\hat{T}}\exp\bigg(-\frac{i}{\hbar}\int^{t_2}_{t_1}\hat{H}(t')\diff t'\bigg),
\end{eqnarray}
where ${\hat{T}}$ is the time-ordering operator \cite{Tannor07}, describes the unitary evolution of the quantum state $\ket{\psi}$ from time $t_1$ to $t_2$. Taking the matrix exponential of the Hamiltonian in the rotating frame (Equation \ref{eq: ham}), it can be shown that \cite{Stoner2011}
\begin{eqnarray}
\hat{U}(t, t+\diff t) = \exp\bigg(-\frac{i {{\mathbf{\Omega}}(t)}\cdot {{\bm{\sigma}}}}{2} \diff t\bigg),
\end{eqnarray}
where ${{\bm{\sigma}}}$ is the vector of Pauli matrices. Therefore, the effect of a pulse of duration $\diff t$ is to rotate the state on the surface of the Bloch sphere by an angle of magnitude  
\begin{eqnarray}
\theta \equiv \tilde{\Omega}_R(t)\diff t = \sqrt{\Omega_R^2 + \delta^2} \times \diff t
\end{eqnarray}
about an axis defined by the field vector ${{\mathbf{\Omega}}(t)}$:
\begin{eqnarray}
{{\mathbf{\Omega}}} = \Omega_R\cos(\phi_L){{\mathbf{x}}} + \Omega_R\sin(\phi_L){{\mathbf{y}}} + (\delta){{\mathbf{z}}}.
\end{eqnarray}
Recalling the following identity, 
\begin{eqnarray}
\exp(i\alpha \mathbf{\hat{n}} \cdot \bm{\sigma}) = \mathbf{I}\cos(\alpha) + \mathbf{\hat{n}} \cdot \bm{\sigma} \sin(\alpha),
\end{eqnarray}
the form of the propagator for a pulse with a constant field vector and duration $\Delta t$ is given by \cite{Stoner2011, Luo2016}
\begin{eqnarray} \label{eq: prop}
\hat{U}  =   \left( \begin{array}{c c}
C^* & -iS^* \\
-iS   &  C
\end{array}
\right)
\end{eqnarray}
where $C$ and $S$ are defined as:
\begin{eqnarray} \nonumber
C &\equiv \cos(\tilde{\Omega}_R\Delta t /2) + i(\delta / \tilde{\Omega}_R)\sin(\tilde{\Omega}_R \Delta t /2)\\
S &\equiv e^{i\phi_L}({\Omega}_R  / \tilde{\Omega}_R)\sin(\tilde{\Omega}_R \Delta t /2).
\end{eqnarray}

\subsection{Inhomogeneities and Composite Pulses}
It is common in NMR spectroscopy to refer to off-resonance and pulse-length errors. An off-resonance error arises when the detuning $\delta$ is non-zero and the field vector does not lie perfectly in the $\vartheta = \pi/2$ plane of the Bloch sphere. A pulse-length error occurs when the desired total rotation angle around the field vector is incorrect, due to either an error in the pulse duration or an error in the effective coupling strength $\tilde{\Omega}_R$. Detunings lead to deflection of the atomic state trajectory, and variations in the coupling strength or Rabi frequency lead to errors in the rotation angle around the field vector. These errors lead to dephasing and reductions in the fidelity of state manipulation \cite{Dunning2014}. 

Composite pulses are pulse sequences intended to replace a desired operation while providing robustness to off-resonance and pulse-length errors. Composite pulses compensate for such errors by using concatenated sequences with tailored phases and durations \cite{Levitt1981}, which are equivalent to series of rotations on the Bloch sphere; both the combined laser phase ($\phi_L$) and pulse duration ($\Delta t$) may be different for each pulse in the sequence. The notation $\theta_{\phi}$ is used to specify a particular rotation element, so that a composite pulse may be written as the sequence: 
\begin{eqnarray}
\theta^{(1)}_{\phi_1}\theta^{(2)}_{\phi_2}\theta^{(3)}_{\phi_3} \ldots.
\end{eqnarray}
Such pulses can be placed into two categories: `point-to-point' (PP) rotations, and `universal rotations' (UR). The first are only designed to perform the correct rotations for specific starting states, whereas UR pulses are intended to perform the correct rotation for all starting points. The type of pulse required depends on the role in the application.

Although numerous composite pulses have been developed for NMR applications, most result from a combination of calculation and intuition and there is no way to design or tailor a sequence for a particular application automatically. Numerical optimization of a control system has become a promising alternative for broadband pulse generation in NMR \cite{Skinner2003}, and has been successfully applied to the control of Bose-Einstein condensates, the stabilization of ultra-cold molecules \cite{VanFrank2014, Jager2014, Koch2004}, and nitrogen-vacancy center magnetometry \cite{Nobauer2015}. Optimal control theory offers methods to generate optimized pulse sequences which are tailored to specific systems and applications.

\section{Optimal Control Theory}
Optimal control theory aims to obtain control parameters that allow a dynamical system to be driven so as to maximize some objective function. In quantum mechanics, this objective function often represents the accuracy with which initial states may be driven to desired final states by the available control fields, subject to constraints on the capabilities of the instruments such as maximum power and frequency. 

The total Hamiltonian of a system may be written in the form
\begin{eqnarray} \label{eq: controlham}
\hat{H}(t) = \hat{H}^{(0)} + \sum_{n=1}^M c^{(n)}(t) \hat{H}^{(n)}
\end{eqnarray}
where the drift Hamiltonian $\hat{H}^{(0)} $  represents the free evolution of the system. The operators $ \hat{H}^{(n)}$ correspond to the experimental control fields whose amplitudes $c^{(n)}(t)$ can be set to form a given composite pulse sequence.

To determine the performance of a given pulse, we consider an example fidelity measure of the following form
\begin{eqnarray}
\mathcal{F}\{c^{(n)}(t)\} =  f\bigg(\bra{\psi_D}{\hat{T}}\bigg[\exp\int^{t_f}_0 -\frac{i}{\hbar}\hat{H}(t) dt \bigg] \ket{\psi_0}\bigg),
\end{eqnarray}
where $\ket{\psi_D}$ is the desired target state, $\ket{\psi_0}$ is the initial state, and $f$ is a differentiable function of the projection of the resultant state onto $\ket{\psi_D}$. The aim is to either minimize or maximize $\mathcal{F}$ for all members of an ensemble with varying drifts. The choice of $\ket{\psi_0}$, $\ket{\psi_D}$, and $f$ is application dependent.

A well-known gradient-based optimization method, first developed by Khaneja \textit{et al.} \cite{Khaneja2005} for the design of NMR pulse sequences, is Gradient Ascent Pulse Engineering (GRAPE). GRAPE has computationally efficient gradients using analytical derivatives \cite{Floether2012, Goodwin2015a}, which can be used to approximate and even compute the Hessian \cite{Goodwin2015} and allow for Newton-Raphson type optimizations.

GRAPE begins by discretizing the M control sequences $c^{(n)}(t)$ into N timesteps $c^{(n)}_k$ of duration $\Delta t$, and assuming that during each timestep the control amplitude $c^{(n)}_k$ is a constant. The optimization involves finding the vectors $\{c^{(n)}_k \}$ such that the chosen functional $\mathcal{F}$ is either maximal or minimal. The form of the time-ordered propagator is then simplified to a product of $k$ ``slices"
\begin{eqnarray}
\hat{U} = \overrightarrow{\prod_k} \exp\bigg(-\frac{i}{\hbar}\big(   \hat{H}^{(0)} + \sum_{n=1}^M c^{(n)}_k \hat{H}^{(n)}    \big)\Delta t_k \bigg) \equiv \overrightarrow{\prod_k} \hat{U}_k.
\end{eqnarray}
Therefore, $\mathcal{F}$ becomes
\begin{eqnarray} \label{eq: J2}
\mathcal{F} = f\bigg(\bra{\psi_D}  \hat{U}_N \hat{U}_{N-1} \ldots \hat{U}_{k} \ldots \hat{U}_{1}   \hat{U}_{0}  \ket{\psi_0} \bigg).
\end{eqnarray}

For the optimization, derivatives of $\mathcal{F}$ are required with respect to the set of control coefficients $\{ c_k^{(n)} \}$. GRAPE computes these derivatives in an efficient way, which can be seen by making the observation that the only element in Equation \ref{eq: J2} that depends on $c^{(n)}_k$ is $\hat{U}_k$. Therefore, ${\partial \mathcal{F}}/{\partial c_k^{(n)}}$ becomes
\begin{eqnarray} 
&\frac{\partial}{\partial c_k^{(n)}}\bigg(\bra{\psi_D}  \hat{U}_N \hat{U}_{N-1} \ldots \hat{U}_{k} \ldots \hat{U}_{1}   \hat{U}_{0}  \ket{\psi_0} \bigg) = \\ 
&\bra{\psi_D}  \hat{U}_N \hat{U}_{N-1} \ldots \hat{U}_{k+1} \frac{\partial \hat{U}_{k}}{\partial c_k^{(n)}} \hat{U}_{k-1}\ldots \hat{U}_{1}   \hat{U}_{0}  \ket{\psi_0}.
\end{eqnarray}
The computation of ${\partial \mathcal{F}}/{\partial c_k^{(n)}}$ for all timesteps $k$ and all control channels $n$ requires just two simulations, namely a backwards propagation from the target state ($\bra{\psi_D}  \hat{U}_N \hat{U}_{N-1} \ldots \hat{U}_{k+1}$) and a forward propagation from the initial state ($\hat{U}_{k-1}\ldots \hat{U}_{1}  \hat{U}_{0}  \ket{\psi_0}$). The directional derivatives which require computation are
\begin{eqnarray}
\frac{\partial}{\partial c_k^{(n)}} \exp\bigg(-\frac{i}{\hbar} \big(   \hat{H}^{(0)} + \sum_{n=1}^M c^{(n)}(t_k) \hat{H}^{(n)}    \big)    \Delta t  \bigg).
\end{eqnarray}

An approximate expression for this derivative is given by Khaneja \textit{et al.} \cite{Khaneja2005}. First-order gradient ascent can then be used to optimize the fidelity $\mathcal{F}$ iteratively. Improvements were made in the use of propagator derivatives by De Fouquieres \textit{et al.} \cite{DeFouquieres2011} which use the gradient to approximate a second order optimization method, called a quasi-Newton optimization, improving optimization convergence. This work uses the limited-memory Broyden-Fletcher-Goldfarb-Shanno (L-BFGS) quasi-Newton method implemented in \textit{Spinach} \cite{DeFouquieres2011, Goodwin2015} with analytical directional derivatives \cite{Goodwin2015a}.

The Hamiltonian of our system (Equation \ref{eq: ham}) can be expressed as
\begin{eqnarray}
\hat{H}_R =  \frac{\delta}{2}\hat{\sigma_z} +\frac{1}{2}\Omega_R(t)\bigg( \cos[\phi_L(t)]\hat{\sigma_x}  +   \sin[\phi_L(t)]\hat{\sigma_y}     \bigg)
\end{eqnarray}
where the drift Hamiltonian is $\hat{H}^{(0)} = (\delta /2)\hat{\sigma_z}$ and the two control operators $\hat{H}^{(1)}$ and $\hat{H}^{(2)}$ can be identified with Pauli matrices $\hat{\sigma_x}$ and $\hat{\sigma_y}$ respectively. The two control coefficients $c^{(1)}(t),c^{(2)}(t)$ are given by $\Omega_R(t) \cos[\phi_L(t)]$ and $\Omega_R(t) \sin[\phi_L(t)]$ respectively. This form is directly analogous to the case of a spin system interacting with an applied rf field given appropriate parameter scaling of the magnitudes of each term \cite{Levitt2008}.

Optimal control can be used to obtain optimal waveforms $\Omega_R(t)$ and $\phi_L(t)$, or the amplitude may be fixed and an optimal phase profile obtained. Initial guesses for the pulse waveform ($\Omega_R(t)$ and $\phi_L(t)$) are provided and the derivatives of the relevant fidelity measure are calculated, returned and fed into the optimization module of \textit{Spinach} \cite{Goodwin2015}. The control pulses are made robust to variations in detuning (off-resonance errors) by providing \textit{Spinach} with an ensemble of drift Hamiltonians and maximizing or minimizing the average of individual fidelities. Robustness to variations in power (pulse-length errors) is achieved by providing a range of power levels, which are averaged over in the fidelity calculation \cite{Kobzar2004, Kobzar2012}. We define the ensemble provided to the optimization as consisting of a number of offsets in detuning and coupling strength, denoted by $\delta^{\mathrm{off}}$ and $\Omega_{R}^{\mathrm{off}}$ respectively. Finally, penalties may be added to the objective function to restrict experimentally relevant quantities such as maximum power or enforce waveform smoothness \cite{Khaneja2005, Goodwin2015}. 

\section{Interferometer Fidelity}

Although mirror pulses may be optimized individually, we obtain measures of mirror pulse fidelity by considering the role of the mirror within a 3-pulse Mach-Zehnder sequence. In addition to the ability of a mirror pulse to transfer as many atoms from ground to excited state as possible, variations in the atomic phase of the final state following a mirror pulse must be minimized to preserve interferometric contrast, as shown by Luo \textit{et al.} \cite{Luo2016}. The excited state population after a sequential application of pulse propagators corresponding to the $\pi/2- \pi- \pi/2$ sequence acting on an atom initially in the ground state can be obtained by following the analytical treatment given by Stoner \textit{et al.} \cite{Stoner2011},
\begin{eqnarray} \label{eq: P1} \nonumber
P_\mathrm{2} &= |S_{\frac{\pi}{2}}|^4|S_{\pi}|^2 + |C_{\frac{\pi}{2}}|^4|S_{\pi}|^2 + 2|S_{\frac{\pi}{2}}|^2|C_{\pi}|^2|C_{\frac{\pi}{2}}|^2  \\
&- 2Re\big(\exp(i\phi_{\mathrm{i}})C_{\frac{\pi}{2}}S_{\frac{\pi}{2}}(S^*_{\pi})^2C^*_{\frac{\pi}{2}}S_{\frac{\pi}{2}}  \big) ,
\end{eqnarray}
where C and S are defined in Equation \ref{eq: prop} and refer to elements of the pulse propagators. We have assumed the initial and final beamsplitters in the Mach-Zehnder sequence to be identical and the dwell times between pulses to be the same. The subscripts `${\frac{\pi}{2}}$' and `${\pi}$' refer to the beamsplitters and mirrors respectively. $\phi_{\mathrm{i}}$ is an interferometric phase term which gives information about the inertial forces acting during the sequence, but which is modified by pulse-dependent phase shifts. This phase must be preserved when averaging over an ensemble of atoms with a distribution of velocities and coupling strengths. 
The interferometer output may be written as \cite{Luo2016}
\begin{eqnarray}\label{eq: target interferometer form}
P_2 = \frac{1}{2} \big( A(\delta) - B(\delta) \cos[\phi_{\mathrm{i}} + \phi_{\mathrm{p}}(\delta)]   \big)
\end{eqnarray}
where $\phi_{\mathrm{p}}(\delta)$ is a phase shift introduced by the pulses and $A(\delta)$ and $B(\delta)$ represent the interferogram offset and contrast respectively. In the ideal case $A(\delta)$, $B(\delta)$, and $\phi_{\mathrm{p}}(\delta)$ should be constant for all detunings present in an atomic sample. For maximum contrast following thermal averaging, A and B must be unity for all atoms. These requirements can be used to obtain a measure of fidelity for the mirror pulse, which we can use to numerically optimize pulse shapes with these properties across an ensemble. The phase shift due to the pulses may be expanded as
\begin{eqnarray}\nonumber
\phi_{\mathrm{p}}(\delta) &=  2\phi S_{\frac{\pi}{2}} - 2\phi S_{\pi} \\
&= 2\phi(i\braket{2|\hat{U}_{\frac{\pi}{2}}|1}) -2\phi(i\braket{2|\hat{U}_{\pi}|1}),
\end{eqnarray}
and the contrast may be written as
\begin{eqnarray} \nonumber
A(\delta) &= 4|C_{\frac{\pi}{2}}|^2 |S_{\pi}|^2 |S_{\frac{\pi}{2}}|^2 \\
&= 4|\braket{1|\hat{U}_{\frac{\pi}{2}}|2}|^2 |\braket{1|\hat{U}_{\pi}|2}|^2 |\braket{1|\hat{U}_{\frac{\pi}{2}}|1}|^2.
\end{eqnarray}
Assuming rectangular and perfectly resonant beamsplitter pulses across the atomic ensemble ($|C_{\frac{\pi}{2}}|^2|S_{\frac{\pi}{2}}|^2 \rightarrow 1/4$) implies that the following constraints on the mirror pulse for an ensemble of atoms will maximize contrast of the Mach-Zehnder output after thermal averaging:
\begin{eqnarray} \label{conditions pi}
\hat{U}_{\pi}\cases{
                  |\braket{1|\hat{U}_{\pi}|1}|^2 = 0    \\
               |\braket{2|\hat{U}_{\pi}|1}|^2  = 1    \\
               \phi(\braket{2|\hat{U}_{\pi}|1})\ \mathrm{constant}.
            }
\end{eqnarray}

We therefore consider the following two mirror pulse fidelities:
\begin{eqnarray} 
\mathcal{F}_{\mathrm{real}}^{\pi} &= Re\braket{2|\hat{U}_{\pi}|1}  \label{UR1}\\ 
\mathcal{F}_{\mathrm{imag}}^{\pi} &= Im\braket{2|\hat{U}_{\pi}|1}  \label{UR2}.
\end{eqnarray}
Maximizing fidelity $\mathcal{F}^{\pi}_{\mathrm{real}}$ or $\mathcal{F}^{\pi}_{\mathrm{imag}}$ constrains the phase of the overlap $\braket{2|\hat{U}_{\pi}|1}$. All the conditions of the optimal inversion pulse (Equations \ref{conditions pi}) are satisfied. Further, it can be shown that these performance functions (Equations \ref{UR1} and  \ref{UR2}) are equivalent to obtaining the universal $180^{\circ}$ rotations explored by Kobzar \textit{et al.} \cite{Kobzar2012}.

Previous work by Dunning \textit{et al.} \cite{Dunning2014} defined the following fidelity for composite mirror pulses:
\begin{eqnarray} \nonumber \label{eq: fid1}
\mathcal{F}_{\mathrm{square}}^{\pi} &=  |\braket{2|\psi_F}|^2 \\
			& = |c_2|^2
\end{eqnarray}
where $c_2$ is the final excited state amplitude of the two-level system and $\ket{\psi_F}$ is the final state. The fidelity in Equation \ref{eq: fid1} does not constrain the phase of the final state and yields pulses which are point-to-point (PP) rotations between the ground and excited states. Since the mirror pulse in the Mach-Zehnder sequence is to be applied to atoms which ideally lie at a range of points on the equator of the Bloch sphere, we expect that pulses which maximize fidelity $\mathcal{F}^{\pi}_{\mathrm{square}}$ will lead to poor interferometric contrast when averaging over an ensemble. Our approach is to use GRAPE to generate pulses which maximize a given mirror fidelity, $\mathcal{F}^{\pi}_{\mathrm{real}}$, $\mathcal{F}^{\pi}_{\mathrm{imag}}$, or $\mathcal{F}^{\pi}_{\mathrm{square}}$ for ensembles with experimentally relevant ranges of detunings and coupling strengths. We then determine through simulation the effect of the cloud temperature alone upon the fringe contrast that these pulses could yield in an interferometric application. 

\section{System and Model Parameters}

Although our approach may be applied to any spin system, we evaluate it for a cold-atom light-pulse interferometer such as that in \cite{Dunning2014}, wherein a thermal cloud of several million atoms of ${}^{85}$Rb at a temperature of order $10-100\ \mu$K is addressed by counter-propagating $780\ $nm laser beams tuned to a Raman transition between the hyperfine levels $\ket{5^2S_{1/2}, F=2}$ and $\ket{5^2S_{1/2}, F=3}$ (levels 1 and 2 respectively). The Raman beams are detuned far from single photon resonance with the intermediate $\ket{5^2 P_{3/2}}$ state, which may theoretically be adiabatically eliminated so that each atom can be described as an effective two-level system \cite{RBerman1997}. A description of our experimental arrangement is given elsewhere \cite{Dunning2014}. Our effective two-state system evolves under the Hamiltonian in Equation \ref{eq: ham}, with $\Omega_R$ replaced by an effective two-photon Rabi frequency ${\Omega_1 \Omega_2}/{(2\Delta)}$, where $\Omega_1$ and $\Omega_2$ represent the coupling of each laser to levels 1 and 2 respectively, and $\Delta$ represents the single photon detuning of both lasers from the intermediate level.  The laser frequency becomes $\omega_L = \omega_{L1}-\omega_{L2}$, where $\omega_{L1,L2}$ are the frequencies of the counter-propagating Raman beams formed by lasers 1 and 2, $\phi_L$ becomes an effective combined laser phase $\phi_{L1} + \phi_{L2}$, and ${\mathbf{k}}_{L}$ is the effective wavevector $ {\mathbf{k}}_{L1} - {\mathbf{k}}_{L2}$.

In our atom cloud, there is Zeeman degeneracy over sub-states distinguished by the quantum number $m_F$, which give rise to multiple coupling strengths, and both the laser intensity and residual magnetic field vary across the atom cloud. The counter-propagating Raman beams give the interaction a Doppler sensitivity, which we use elsewhere for velocimetry and inertial measurement \cite{Carey2017}. Different atoms thus see different coupling strengths and Doppler, Zeeman and AC Stark shifts according to their internal and external states. Both pulse-length and off-resonance errors are present in our system. For an experimentally-measured Rabi frequency of $\Omega_{\mathrm{eff}} \approx 2\pi \times 360\ $kHz we find coupling strength variations of approximately $0.3\  \Omega_{\mathrm{eff}}$ and off-resonance errors due to a Gaussian velocity distribution with a full width at half maximum of approximately $1.5\ \Omega_{\mathrm{eff}}$ in a $\sim 80\ \mu$K cloud \cite{Dunning2014a}.

\section{Results \label{Results}}

All pulse optimizations were carried out using GRAPE and \textit{Spinach}, constraining the effective Rabi frequency to correspond to a limited laser power and fixed single-photon detuning, and with a discretization timestep of 100 ns. Optimizations were carried out for ensembles with various ranges of detunings and coupling strengths (see Table \ref{TABLE1}), using fidelities $\mathcal{F}^{\pi}_{\mathrm{real}}$, $\mathcal{F}^{\pi}_{\mathrm{imag}}$, and $\mathcal{F}^{\pi}_{\mathrm{square}}$. Longer pulses can excite atoms from the ground to excited states for larger detunings than shorter pulses. Mirror pulses are compared based on their ability to transfer atoms from state to state and the magnitude of the atomic phase variations over the simulated ensemble of atoms. The mirror pulse is characterized by a time-varying phase, initially set to be flat, and optimization was continued until convergence. Different initial pulse shapes were often found to lead to the same optimal pulse, and pulses often exhibited symmetry about the mid-point - a phenomenon also seen in NMR \cite{Kobzar2004, Kobzar2008a, Kobzar2012}. We begin by presenting the results obtained by considering a general ensemble of pulse-length and off-resonance errors in our optimization, before applying our chosen optimal control method to an ensemble representing the experimental inhomogeneities present in a particular Raman polarization arrangement.

\subsection{Mirror pulse optimization}

\begin{figure}[!htb]
\centering
   \includegraphics[width=0.7\linewidth]{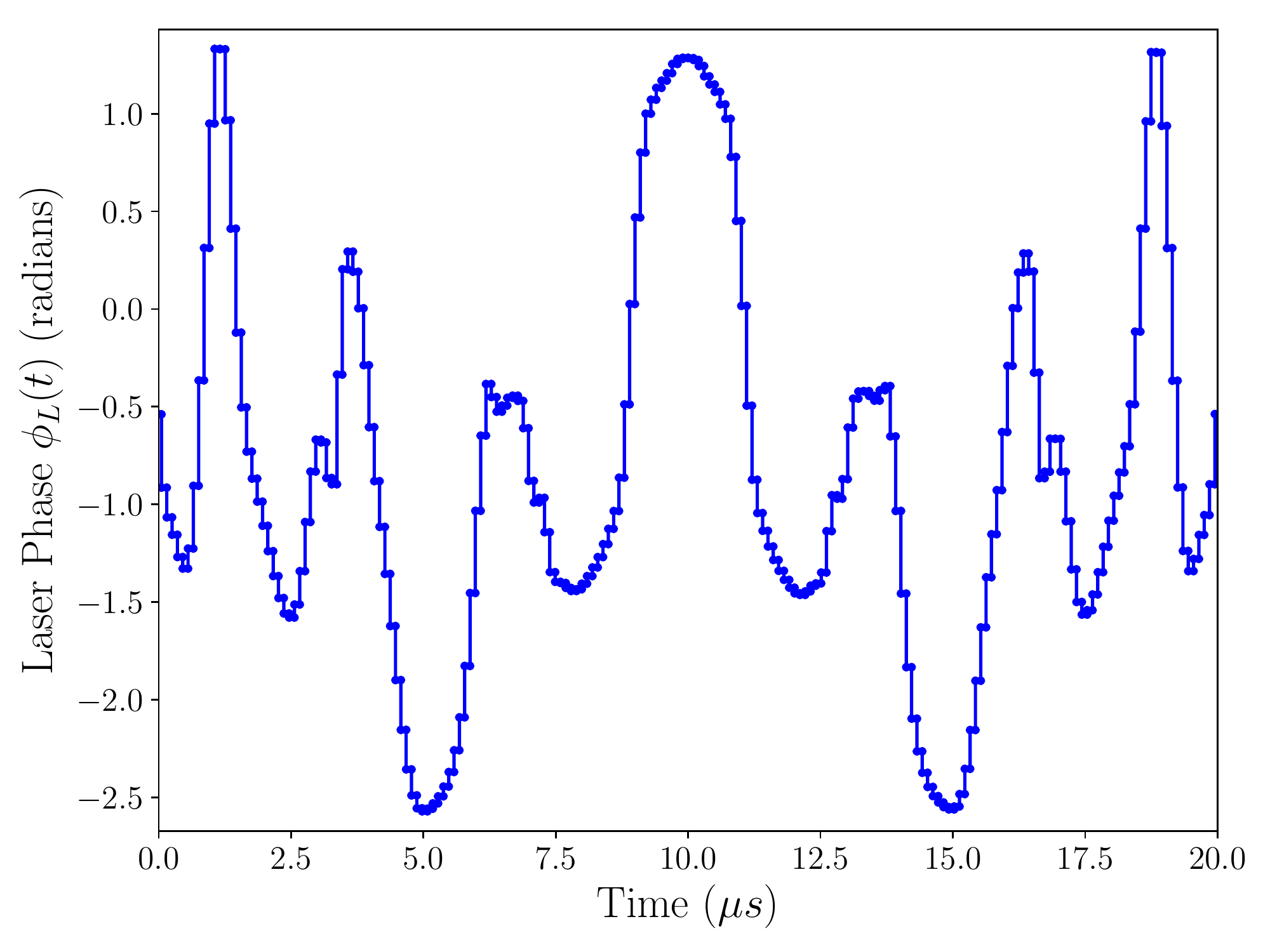}
\caption{Phase profile $\phi_L(t)$ for GRAPE inversion pulse optimizing $\mathcal{F}^{\pi}_{\mathrm{real}}$ subject to constrained total duration of 20 $\mu$s and fixed effective Rabi frequency of $2\pi \times 200\ $kHz. The fidelity after 100 iterations was 0.99.}
\label{MIRROR1}
\end{figure}

Figure \ref{MIRROR1} shows the phase profile of a constant-amplitude mirror pulse obtained by maximizing $\mathcal{F}^{\pi}_{\mathrm{real}}$ subject to a constrained duration of 20 $\mu$s, for a detuning range of $\pm 1.5\ \Omega_{\mathrm{eff}}$ and a power range of $\pm 0.1$ in $\Omega_{\mathrm{eff}}$. If the duration of the pulse and the ensemble detuning range are increased, further variations in phase appear symmetrically around the centre of the pulse shape. In practice, a balance must be struck between pulse duration and performance, dependent on the required application and capabilities of the experimental apparatus to produce complex waveforms. Table \ref{TABLE1} summarises the performance of composite mirror pulses and GRAPE optimizations carried out for different ensembles and fidelities.

\begin{figure}[!htb]
\centering
   \includegraphics[width=0.7\linewidth]{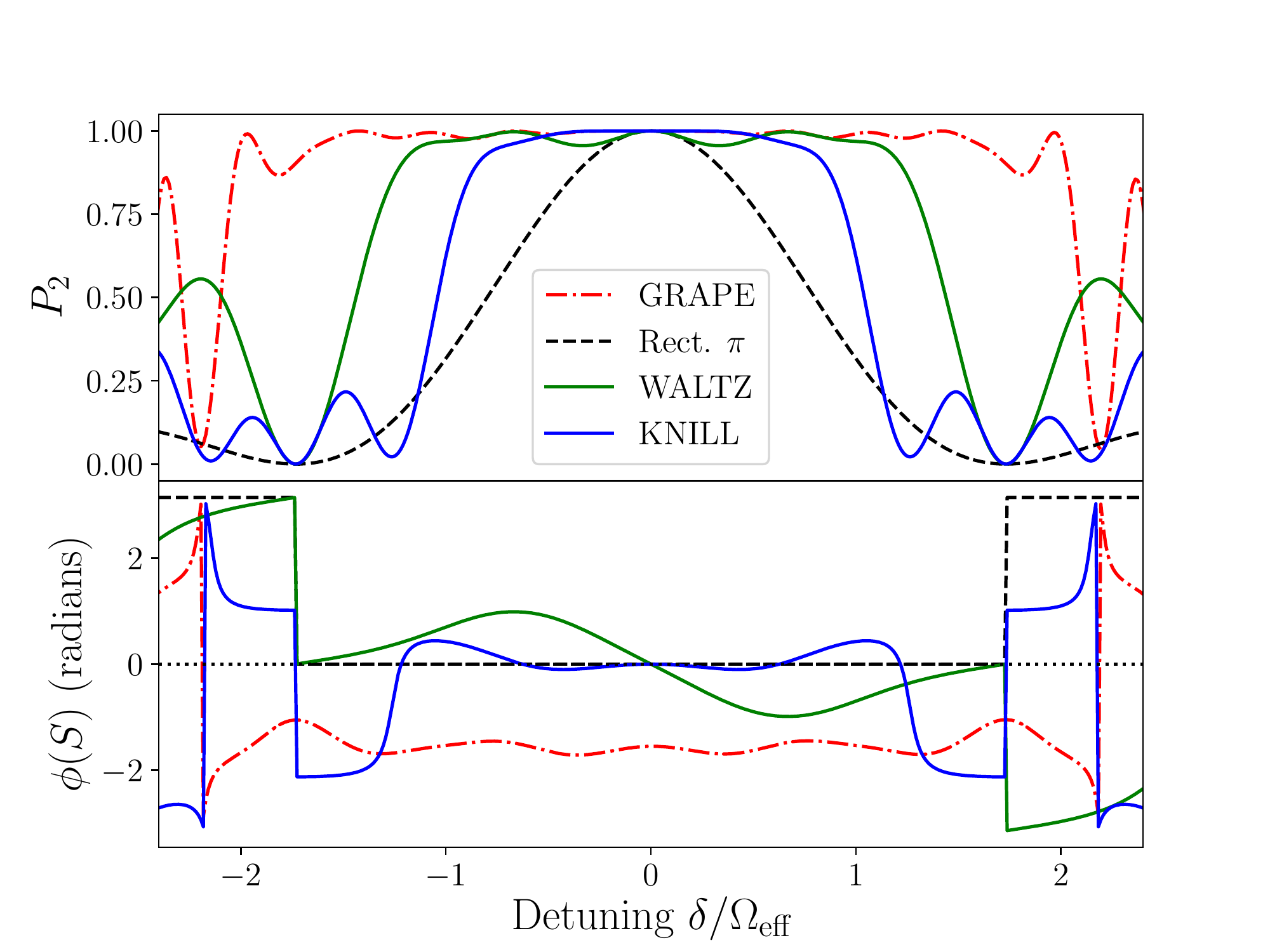}
\caption{Excited state probability and phase profiles plotted against detuning for various composite inversion pulses and GRAPE pulse of Figure \ref{MIRROR1}. The simulation assumes a single Rabi frequency of $2\pi \times 200\ $kHz.}
\label{MIRROR1_prob/phase}
\end{figure}

The response purely as a function of the detuning from resonance is shown in Figure \ref{MIRROR1_prob/phase}. The detuning here would be experimentally due to Doppler and Zeeman shifts. Both the excited state probability after application of the pulse to atoms in the ground state, and the phase response of the $S$ element of the propagator are shown. GRAPE maintains a transition probability $>$0.976 over the optimized range of offsets ($\pm1.5\ \Omega_{\mathrm{eff}}$). For comparison, we show the corresponding results for the rectangular $\pi$ pulse and for WALTZ and KNILL pulses \cite{Shaka1983, Ryan2010} of the same intensity. The WALTZ pulse is highly robust to detunings, but is designed as a ‘point-to-point’ operation for a particular starting state and hence shows large variations in the atomic phase as the detuning is varied; the GRAPE pulse, in contrast, shows smaller phase variations over the range of detunings. The conventional $\pi$ pulse shows an ideal, flat phase response but very narrow range of detunings over which it is efficient.
 
\begin{figure}[!htb]
\centering
   \includegraphics[width=0.7\linewidth]{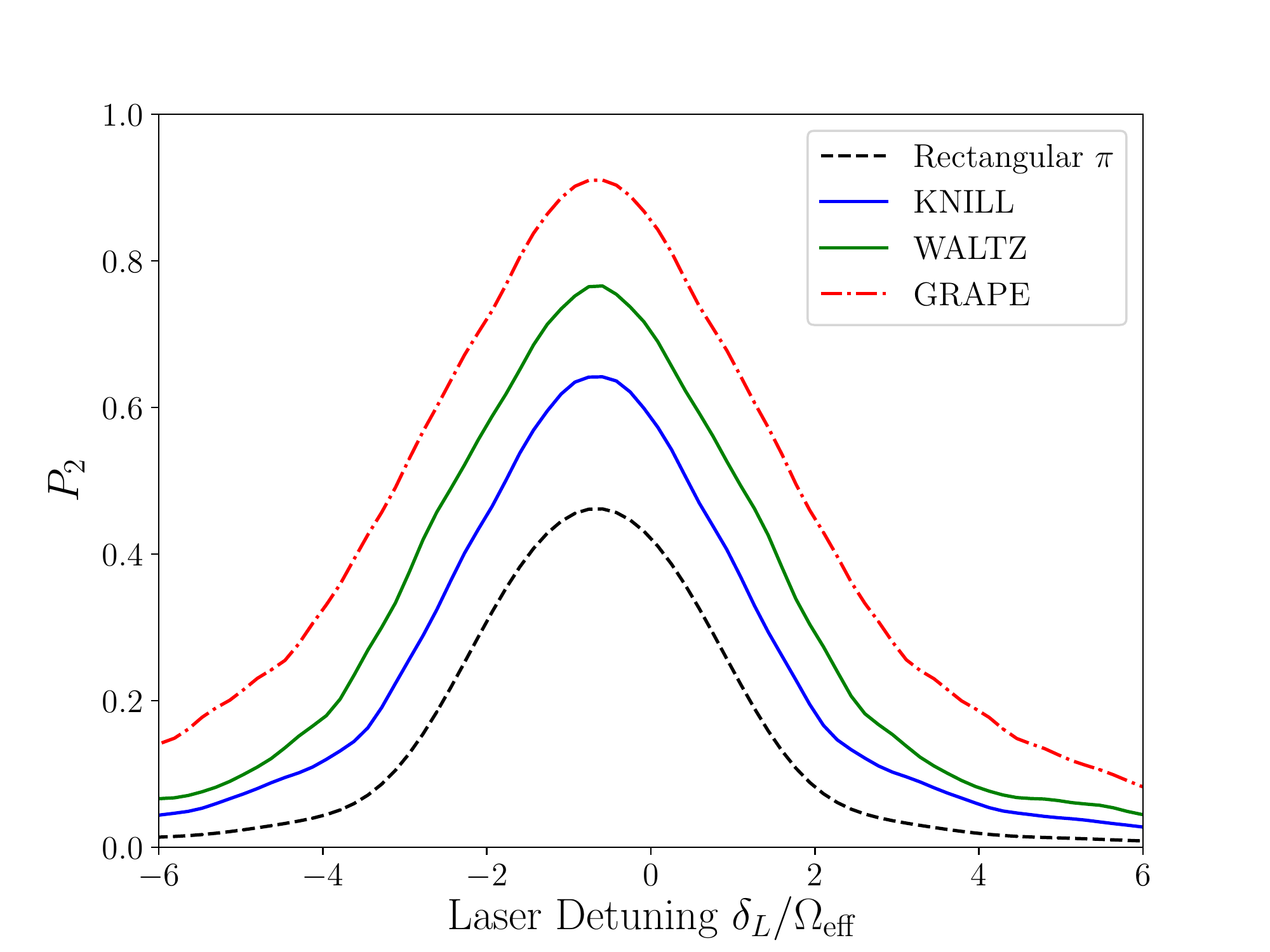}
\caption{Simulation of inversion operation for multiple composite pulses and the GRAPE pulse of Figure \ref{MIRROR1} in a $\sigma^{+}-\sigma^{+}$ Raman arrangement, as the laser detuning $\delta_L$ (defined as $\omega_{12} - \omega_L$) is scanned. This simulation uses an effective Rabi frequency of $2\pi \times 360\ $ kHz, and accounts for both the coupling strengths of 5 $m_{\mathrm{F}}$ sublevels and the Stark shift. Both the simulation parameters and model are described elsewhere \cite{Dunning2014, Dunning2014a}. The offset in peak is due to the Stark shift, and the magnitude of the peak gives an indication of a pulse's ability to excite atoms across the momentum distribution for all $m_{F}$ sublevels, which we assume are equally populated.}
\label{MIRROR_dunning}
\end{figure}

\begin{figure*}[!htb]
\subfloat[Rectangular $\pi$ pulse.]{\includegraphics[width=.49\textwidth]{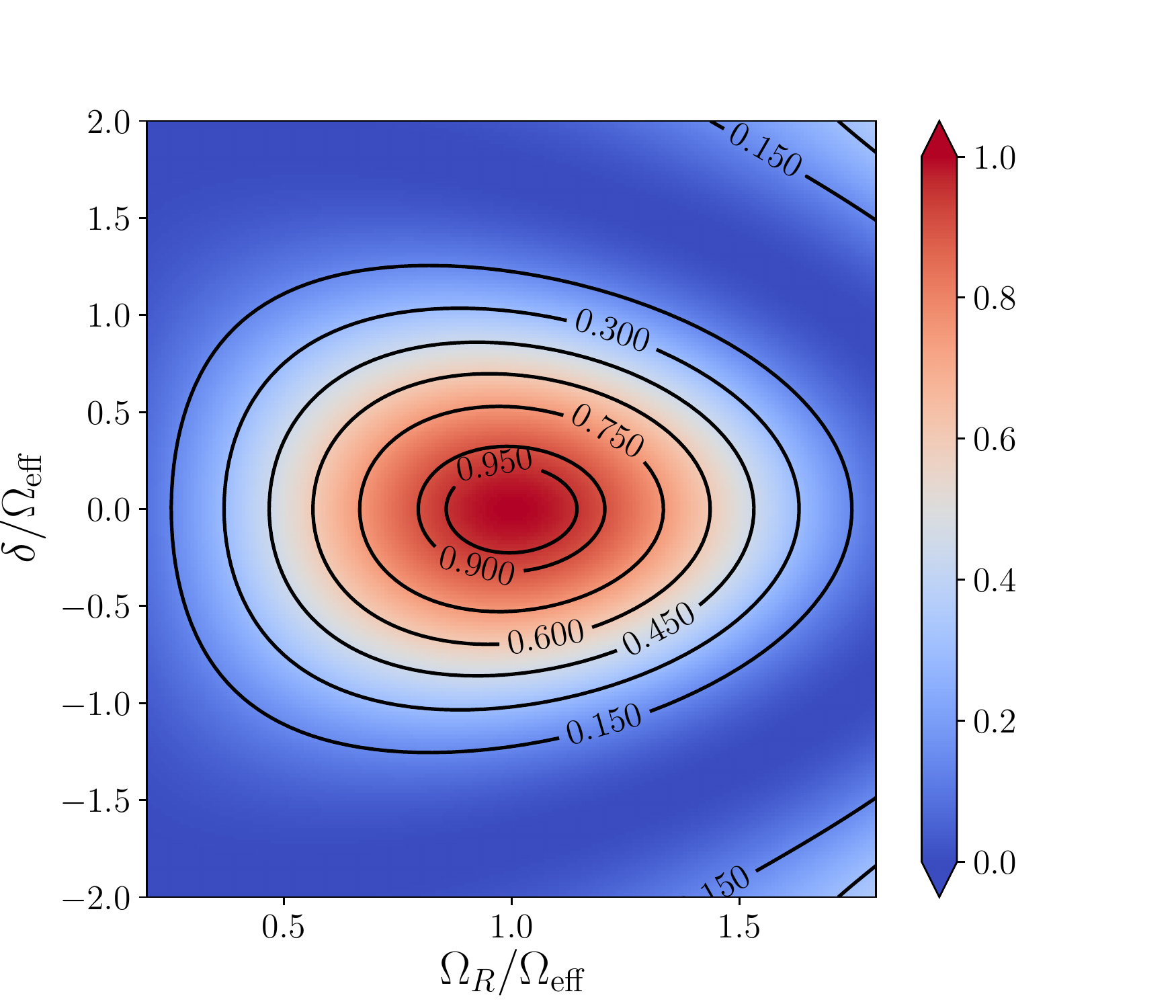}}\quad
\subfloat[KNILL pulse.]{\includegraphics[width=.49\textwidth]{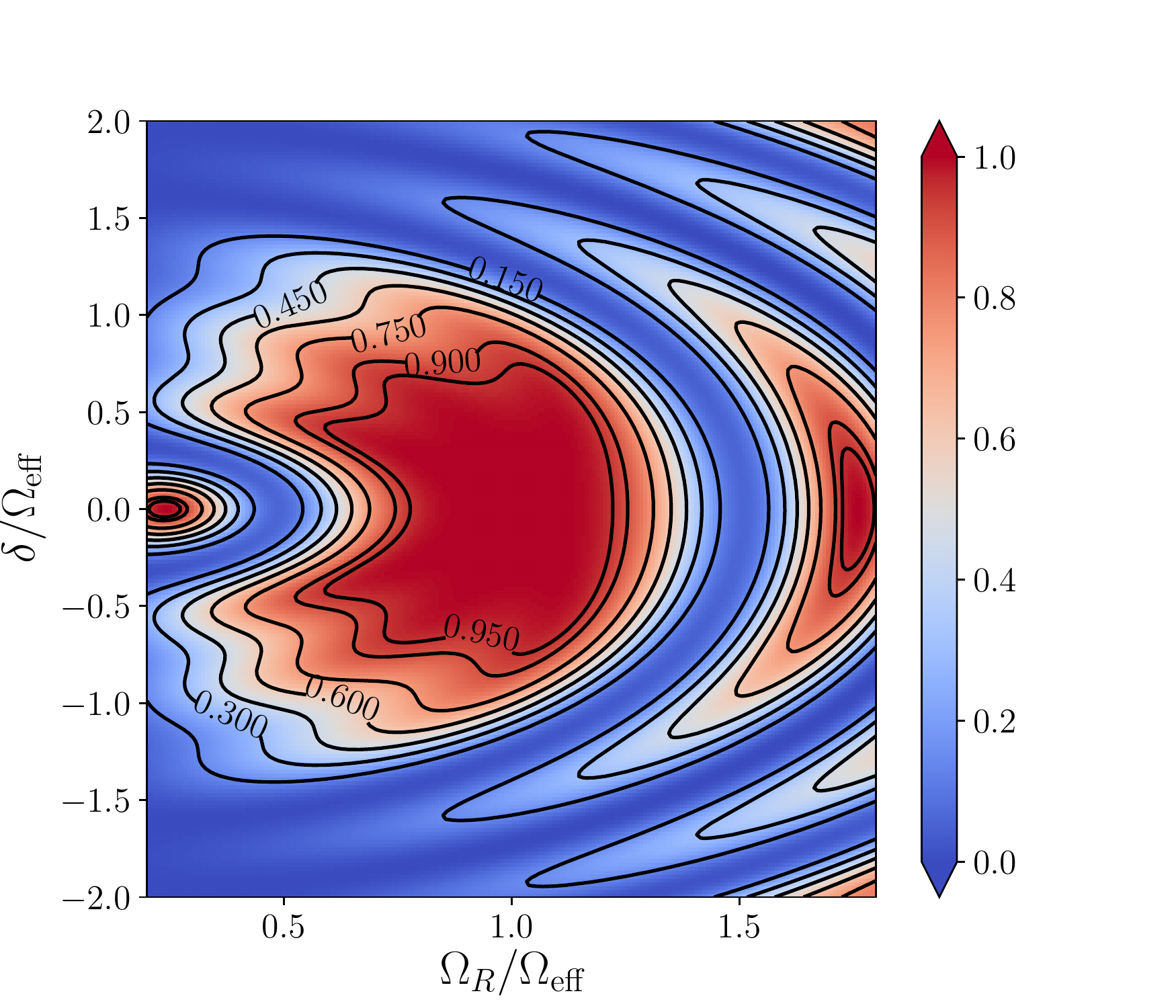}}\\
\subfloat[WALTZ pulse.]{\includegraphics[width=.49\textwidth]{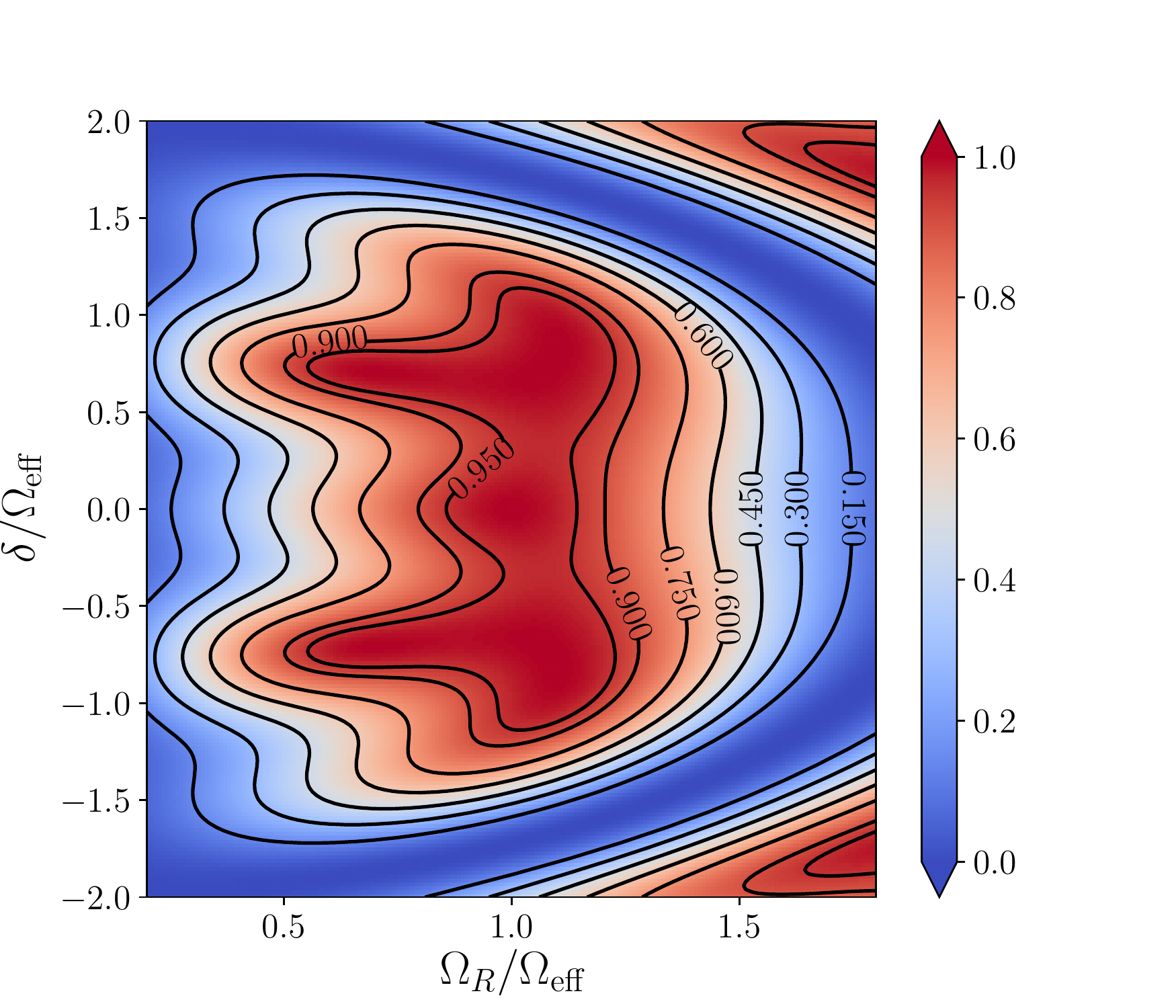}}\quad
\subfloat[GRAPE pulse from Figure \ref{MIRROR1}.]{\includegraphics[width=.49\textwidth]{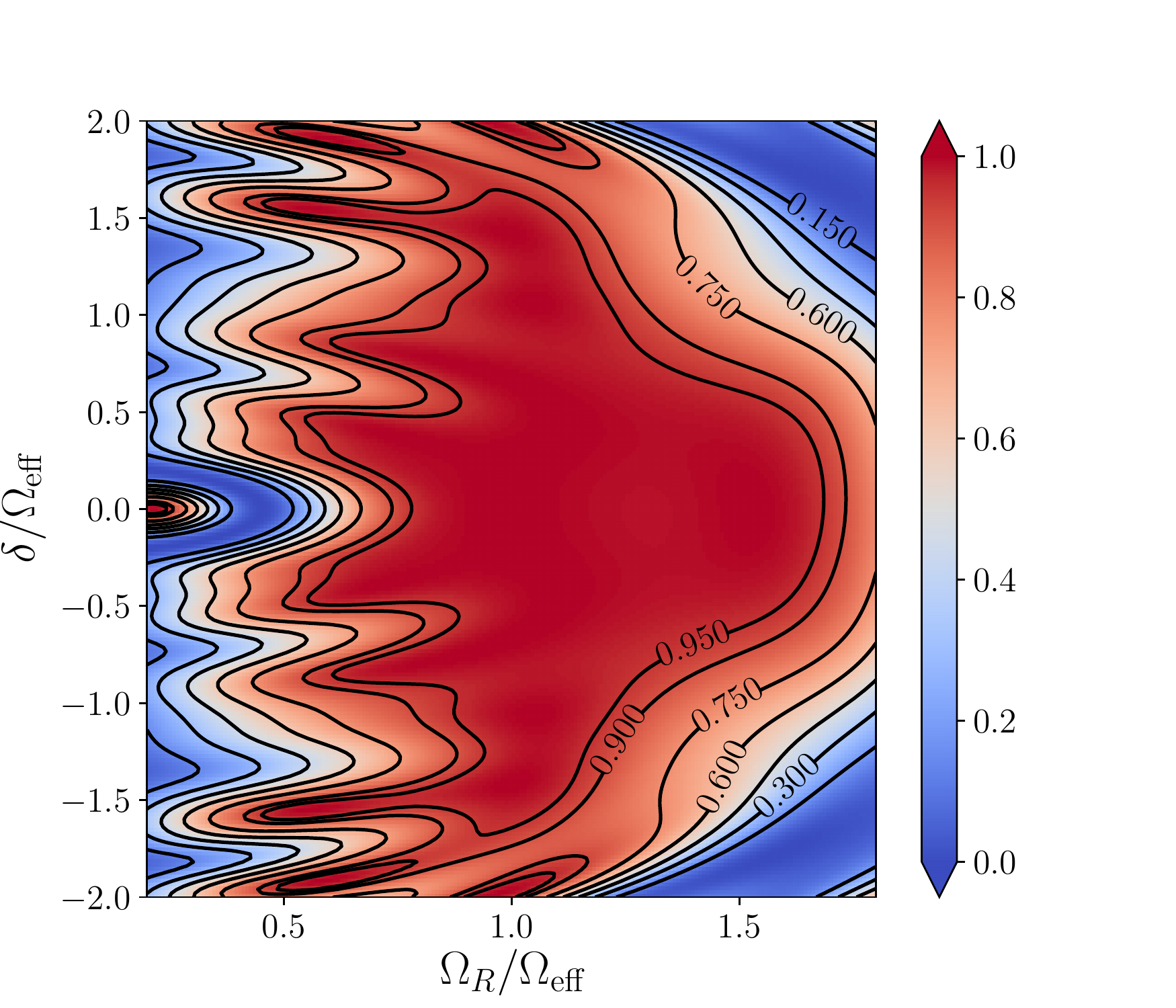}}
\caption{Final excited state population shown for a range of off-resonance and pulse-length errors for different mirror pulses. The simulations were performed at the same effective Rabi frequency of $2\pi \times 200\ $kHz. The GRAPE mirror pulse demonstrates a clear improvement in robustness to variations in detuning and pulse amplitude. The contours are at 0.15, 0.3, 0.45, 0.6, 0.75, 0.9, and 0.95.}
\label{fig:contour1}
\end{figure*}

GRAPE can be used to optimize a mirror pulse for an experimental range of off-resonance errors and the pulse-length errors which arise from the degenerate Zeeman sublevels in the $\sigma^{+}-\sigma^{+}$ Raman polarization arrangement. We simulate the pulse profile of Figure \ref{MIRROR1} using a model which accounts for the AC Stark shift, $m_F$ levels, measured atomic momentum distribution, and experimental parameters in the $\sigma^{+}-\sigma^{+}$ polarization arrangement for a cloud with temperature $\sim 80\ \mu$K (see Figure \ref{MIRROR_dunning}). Such an arrangement has large off-resonance errors and is therefore a good choice to compare pulse performance. The numerical model used in this simulation was shown to have good agreement with experiment \cite{Dunning2014, Dunning2014a}. In this simulation, the final excited state population is calculated numerically and averaged over a range of atomic momenta from the measured distribution, and the five Zeeman sublevels. The GRAPE simulations show an improvement in peak fidelity by a factor of $1.2$ compared with WALTZ, and a factor of $2$ compared with the rectangular $\pi$ pulse. As the laser detuning is scanned, the peak excited state population depends on how well a mirror pulse can excite atoms from the ground to excited state for the distribution of coupling strengths and velocity classes in the ensemble. Figure \ref{MIRROR_dunning} demonstrates GRAPE's ability to compensate for the off-resonance and pulse-length errors present in this system. Contour plots (Figure \ref{fig:contour1}) demonstrate the superior robustness to both error classes offered by our chosen optimal control method.

\subsection{Simulated interferometric contrast}
The temperature-dependent contrast resulting from the Mach-Zehnder interferometer sequence was determined following the procedure outlined by Luo \textit{et al.} \cite{Luo2016} and integrating over Maxwell-Boltzmann velocity distributions with temperatures in the range ($10^{-1}- 10^{3}\ \mu$K). A uniform pulse intensity and single coupling strength, with no $m_F$-dependent inhomogeneity, were assumed throughout. The beamsplitter and recombiner pulses were taken to be rectangular $\pi/2$ pulses with a fixed effective Rabi frequency of $2\pi \times 200\ $kHz, and hence limit the achievable fringe contrast as shown by their combination with a ‘perfect $\pi$’ mirror pulse that is taken to perform an ideal rotation for all atoms. With a realistic mirror formed by a rectangular $\pi$ pulse with the same Rabi frequency as the beamsplitter and recombiner, the interferometer contrast is limited by the imperfections in the rectangular $\pi$ pulse to around 0.8 at a temperature of 20$\mu$K.

The GRAPE pulse of Figure \ref{MIRROR1} offers an improvement in fringe contrast by a factor of 1.76 over that with the rectangular $\pi$ pulse at 100$\mu$K, and approaches the fidelity predicted with a perfect $\pi$ pulse. This limit should be approached more closely as the GRAPE pulse duration is increased, as longer pulses can compensate for a greater range of off-resonance errors. As noted by Luo \textit{et al.} \cite{Luo2016, Luo2016a}, the WALTZ pulse performs badly due to its non-uniform phase response, as it was designed as a point-to-point pulse whereas the mirror operation must have high fidelity irrespective of the starting point on the equator of the Bloch sphere. The point-to-point $\mathcal{F}_{\mathrm{square}}^{\pi}$ GRAPE pulse also performs poorly, as expected. We find the KNILL pulse, designed as a universal $180^{\circ}$ rotation, offers a slight improvement over the sequence of rectangular pulses for the simulated range of temperatures, but is outperformed by GRAPE at all temperatures. 
                                                
\begin{figure}[!htb]
\centering
   \includegraphics[width=0.7\linewidth]{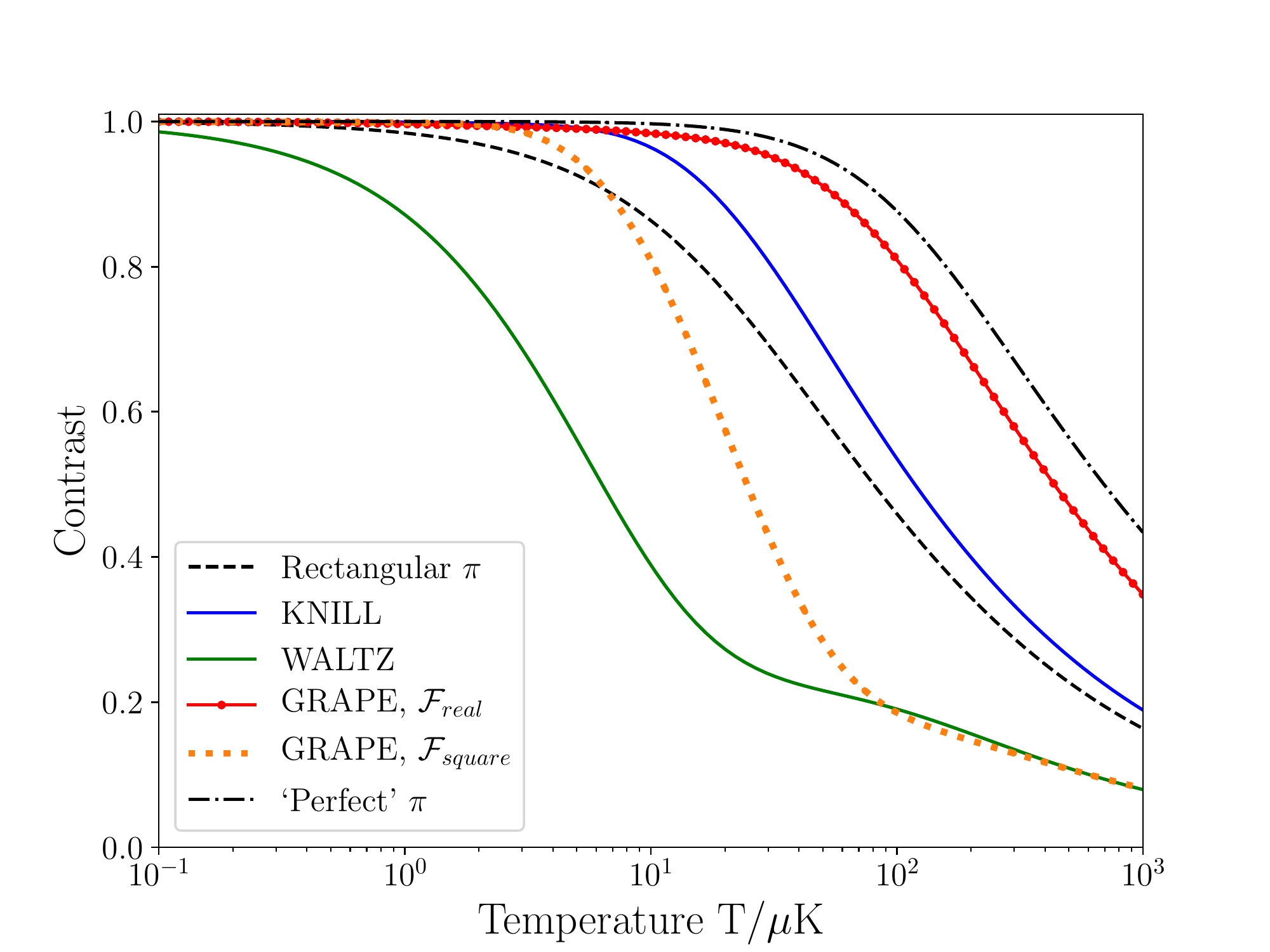}
\caption{Contrast after Mach-Zehnder sequence integrated over thermal distributions of ${}^{85}$Rb atoms with various temperatures. The beamsplitters in each case are rectangular $\pi/2$ pulses. There is a large increase in contrast due to the GRAPE UR mirror pulse in Figure \ref{MIRROR1} which improves the rectangular $\pi$ pulse by a factor of 1.76 at $100\ \mu $K, approaching the theoretical limit of a sequence with a `perfect' $\pi$ pulse resonant for all atoms in the sample. Comparisons with different sequences are made at same Rabi frequency of $2\pi \times 200\ $kHz.}
\label{CONTRAST1}
\end{figure}

\section{Discussion}

Tolerance of ‘pulse-length’ and ‘off-resonance’ errors is essential for the pulse operations in atom interferometers, where a range of velocities, beam intensities and Zeeman substates may be encountered. We have used the optimal control technique of gradient ascent pulse engineering (GRAPE) to obtain robust ‘mirror’ pulses, tailored to accommodate the inhomogeneities found in cold-atom matterwave interferometers, and find such pulses to outperform all the other composite pulses that we have tested. By using a numerical model which has been shown to agree well with experiment \cite{Dunning2014}, we have simulated the performance of an atom interferometer for the $\sigma^{+}-\sigma^{+}$ polarization Raman arrangement within a $\sim 80\ \mu$K cloud of atomic rubidium subject to realistic intensity inhomogeneities and Zeeman substate distributions, and find that our GRAPE pulses show a peak fidelity twice that obtained with a basic rectangular $\pi$ pulse, and 1.2 times that achieved using the WALTZ sequence, with significantly greater tolerance of variations in the atom velocity.

The improved fidelity should allow improvements in the sensitivity of interferometric measurements by permitting greater use of augmentation pulses for large momentum transfer interferometers \cite{Butts2013}, while the tolerance to atom velocity variations will lower measurement noise by allowing the use of warmer atom clouds, and hence higher atom numbers, without incurring the losses of further cooling or filtering. Our GRAPE pulses should provide, for example, transfer efficiencies above 0.9 for a detuning range 1.4 times that tolerated by the WALTZ sequence, which was otherwise the best pulse tested for this system \cite{Dunning2014}.

Replacing the basic rectangular $\pi$ pulse with a GRAPE mirror in a Mach-Zehnder arrangement at 100$\mu$K is sufficient to improve the simulated interferometer contrast by a factor of 1.76, or to achieve the same contrast as a basic $\pi$ pulse for a 15$\mu$K atom cloud. At higher atom temperatures, we see that the interferometer fidelity is limited principally by the fidelity of the beamsplitter operations.

Our optimal control approach depends upon an appropriate choice of the measure of performance. We find that those used for broadband UR $180^{\circ}$ pulses, such as $\mathcal{F}^{\pi}_{\mathrm{real}}$ (Equation \ref{UR1}) and equivalents considered by Kobzar \textit{et al}. \cite{Kobzar2012}, are able to preserve the interferometric phase and increase contrast, and when our optimization is carried out for small detuning ranges we produce similar pulse shapes. Measures of performance more suitable for point-to-point operations, such as $\mathcal{F}_{\mathrm{square}}^{\pi}$ (Equation \ref{eq: fid1}), conversely yield lower interferometer contrast, as does the WALTZ point-to-point composite pulse.

While GRAPE should also be applicable to the design of beamsplitter pulses, we expect that a more fruitful approach will be to consider symmetries in the Mach-Zehnder sequence and compensate in the second beamsplitter operation for errors introduced in the first, so as to optimize the interferometer as a whole. Such cooperative pulse optimization was investigated by Braun \textit{et al}. \cite{Braun2010, Braun2014} for Ramsey-style experiments in NMR, and allows greater freedom in the optimization as individual beamsplitter pulses are permitted imperfections provided that they are cancelled elsewhere in the interferometer sequence. This should allow shorter pulse sequences, desirable for interferometric sensors operating in dynamic environments \cite{Stoner2011} and is attractive for optimization of the $\pi/2-\pi-\pi/2$ sequence used for inertial sensing applications. Nonetheless, we expect the mirror optimization described here to suffice for a large contrast improvement in many current configurations.

Our future work will involve experimental demonstration of GRAPE mirror pulses in our experiment, and analysis of how features present in pulse profiles from optimal control methods allow for error compensation. This will involve characterizing the dynamics and evolution of atomic trajectories on the Bloch sphere.
                              
\begin{center}{
\begin{table*}[!htb]
		\caption{Performance of GRAPE pulses compared with composite mirror pulses. Robustness is measured in units of the effective Rabi frequency $\Omega_{\mathrm{eff}}$ and represents the range of detuning for which the final excited state probability is $>$0.5 and $>$0.9 after application to atoms in the ground state. The maximum phase response variation, $\Delta \phi(S)$,  is taken over a range of $\pm\  \Omega_{\mathrm{eff}}$ in $\delta$. Optimization parameters are provided for GRAPE pulses including the offsets optimized for in pulse-length and off-resonance errors, and the number of offsets used. In each optimization, ensembles were weighted by 8 additional detuning offsets near resonance. The best values for the standard composite sequences tested and GRAPE sequences are in bold.}
\label{TABLE1}
		\footnotesize{
		\begin{tabular}{@{}lllllll }
		\br
			  & Length & \multicolumn{2}{l}{Sequence  }& 		\multicolumn{2}{l}{Robustness ($\delta/\Omega_{\mathrm{eff}}$)}&Max $\Delta \phi(S)$\\
			Pulse& ($t/t_{\pi}$)& $\theta^{(1)}_{\phi_1}\theta^{(2)}_{\phi_2}\theta^{(3)}_{\phi_3} \ldots.$ & & $>0.5$  & $>0.9$ & (radians)  \\
			\mr
			Rectangular $\pi$   & 1       & \multicolumn{2}{l}{$180_{0}$} &   1.597 & 0.645  &  0 \\
			Levitt  \cite{Levitt1981a}      &   2     &  \multicolumn{2}{l}{$90_{90}180_{0}90_{90} $}  &  2.637 & 2.112  & 0.953 \\
			WALTZ \cite{Shaka1983}         & 3       &  \multicolumn{2}{l}{$90_{0}180_{180}270_{0}$}& \textbf{2.878}   &  \textbf{2.434} &  1.974 \\
			KNILL \cite{Ryan2010}   & 5       &  \multicolumn{2}{l}{$180_{240}180_{210}180_{300}180_{210}180_{240} $}& 2.082   &1.693 & \textbf{0.528} \\
			CORPSE \cite{Cummins2000}       &   4.333  & \multicolumn{2}{l}{$60_{0}300_{180}420_{0} $}&      1.438     &1.004 & 2.717 \\
			SCROFULOUS \cite{Cummins2003}  & 3       & \multicolumn{2}{l}{$180_{60}180_{300}180_{60}$} &1.347 & 0.334&  0.834\\
			BB1 \cite{Wimperis1994}    & 5       & \multicolumn{2}{l}{$180_{104.5}360_{313.4}180_{104.5}180_{0}$} &1.685 & 1.106 & 1.778  \\
			\mr
		       &    & \multicolumn{2}{l}{Ensemble (number, range)} &         & &  \\
		      			GRAPE fidelity& & $\delta^{\mathrm{off}}$ & ${\Omega_R}^{\mathrm{off}}$  &  &  & \\

		    \mr
			$\mathcal{F}_{\mathrm{real}}^{\pi} $ fig. \ref{MIRROR1}      & 8       & $20$, $\pm 1.5\Omega_{\mathrm{eff}}$ &$5$, $\pm 0.1\Omega_{\mathrm{eff}}$ & 4.194 & 3.470 & 0.269 \\
			$\mathcal{F}_{\mathrm{square}}^{\pi} $   & 8       & $20$, $\pm 1.5\Omega_{\mathrm{eff}}$ &$5$, $\pm 0.1\Omega_{\mathrm{eff}}$                            &  3.968 & 3.376 & 1.562 \\
			$\mathcal{F}_{\mathrm{imag}}^{\pi} $    & 8         &$20$, $\pm 1.5\Omega_{\mathrm{eff}}$& $5$, $\pm 0.1\Omega_{\mathrm{eff}}$                             & 3.904  & 3.259 & 0.137 \\
			$\mathcal{F}_{\mathrm{real}}^{\pi} $      & 16     & $30$, $\pm 2\Omega_{\mathrm{eff}}$&$5$, $\pm 0.1\Omega_{\mathrm{eff}}$                                  & 4.302 & 4.128 & \textbf{0.096}\\
			$\mathcal{F}_{\mathrm{real}}^{\pi} $  & 32            & $40$, $\pm 2.5\Omega_{\mathrm{eff}}$& $5$, $\pm 0.1\Omega_{\mathrm{eff}}$                           & \textbf{5.513} & \textbf{5.109}& 0.216 \\
			\br
		\end{tabular}
	  	}
\end{table*}
}
\end{center}

\ack{
The \textit{Spinach} spin dynamics software suite and its optimal control module were used \cite{Hogben2011}. The authors are grateful for the advice provided by David Elcock and Matt Himsworth. This work was supported by the EPSRC through the UK Quantum Technology Hub for Sensors $\&$ Metrology under grant EP/M013294/1, the Centre for Doctoral Training in Next Generation Computational Modelling under grant EP/L015382/, and by Dstl under grants DSTLX-1000091758 and DSTLX-1000097855.
}

\section*{References}
\bibliographystyle{unsrt}
\bibliography{optimalcontrolsaywell.bib}

\begin{thebibliography}{10}

\bibitem{Kasevich1991}
Mark Kasevich and Steven Chu.
\newblock {Atomic interferometry using stimulated Raman transitions}.
\newblock {\em Physical Review Letters}, 67(2):181--184, jul 1991.

\bibitem{Gustavson1997}
T.~L. Gustavson, P.~Bouyer, and M.~A. Kasevich.
\newblock {Precision rotation measurements with an atom interferometer
  gyroscope}.
\newblock {\em Physical Review Letters}, 78(11):2046--2049, mar 1997.

\bibitem{DeAngelis2009}
M.~{De Angelis}, A.~Bertoldi, L.~Cacciapuoti, A.~Giorgini, G.~Lamporesi,
  M.~Prevedelli, G.~Saccorotti, F.~Sorrentino, and G.~M. Tino.
\newblock {Precision gravimetry with atomic sensors}.
\newblock {\em Measurement Science and Technology}, 20(2):022001, feb 2009.

\bibitem{Barrett2014}
Brynle Barrett, R{\'{e}}my Geiger, Indranil Dutta, Matthieu Meunier, Benjamin
  Canuel, Alexandre Gauguet, Philippe Bouyer, and Arnaud Landragin.
\newblock {L'effet Sagnac: 20 ans de d{\'{e}}veloppements des
  interf{\'{e}}rom{\`{e}}tres {\`{a}} ondes de mati{\`{e}}re}.
\newblock {\em Comptes Rendus Physique}, 15(10):875--883, dec 2014.

\bibitem{RBerman1997}
Paul Berman, editor.
\newblock {\em {Atom Interferometry}}.
\newblock Academic Press, 1997.

\bibitem{McGuirk2000}
J.~M. McGuirk, M.~J. Snadden, and M.~A. Kasevich.
\newblock {Large area light-pulse atom interferometry}.
\newblock {\em Physical Review Letters}, 85(21):4498--4501, nov 2000.

\bibitem{Islam2008}
Rameez~Ul Islam, Ashfaq~H. Khosa, and Farhan Saif.
\newblock {Generation of Bell, NOON and W states via atom interferometry}.
\newblock {\em Journal of Physics B: Atomic, Molecular and Optical Physics},
  41(3):035505, feb 2008.

\bibitem{Dunning2014}
Alexander Dunning, Rachel Gregory, James Bateman, Nathan Cooper, Matthew
  Himsworth, Jonathan~A. Jones, and Tim Freegarde.
\newblock {Composite pulses for interferometry in a thermal cold atom cloud}.
\newblock {\em Physical Review A - Atomic, Molecular, and Optical Physics},
  90(3):033608, sep 2014.

\bibitem{Kasevich1991a}
Mark Kasevich, David~S. Weiss, Erling Riis, Kathryn Moler, Steven Kasapi, and
  Steven Chu.
\newblock {Atomic velocity selection using stimulated Raman transitions}.
\newblock {\em Physical Review Letters}, 66(18):2297--2300, may 1991.

\bibitem{McGuirk2002}
J.~M. McGuirk, G.~T. Foster, J.~B. Fixler, M.~J. Snadden, and M.~A. Kasevich.
\newblock {Sensitive absolute-gravity gradiometry using atom interferometry}.
\newblock {\em Physical Review A}, 65(3):033608, feb 2002.

\bibitem{Szigeti2012a}
S~S Szigeti, J~E Debs, J~J Hope, N~P Robins, and J~D Close.
\newblock {Why momentum width matters for atom interferometry with Bragg
  pulses}.
\newblock {\em New Journal of Physics}, 14(2):023009, feb 2012.

\bibitem{Butts2013}
David~L Butts, Krish Kotru, Joseph~M Kinast, Antonije~M Radojevic, Brian~P
  Timmons, and Richard~E Stoner.
\newblock {Efficient broadband Raman pulses for large-area atom
  interferometry}.
\newblock {\em Journal of the Optical Society of America B}, 30(4):922, apr
  2013.

\bibitem{Freeman1998}
Ray Freeman.
\newblock {Shaped radiofrequency pulses in high resolution NMR}.
\newblock {\em Progress in Nuclear Magnetic Resonance Spectroscopy},
  32(1):59--106, feb 1998.

\bibitem{Luo2016}
Yukun Luo, Shuhua Yan, Qingqing Hu, Aiai Jia, Chunhua Wei, and Jun Yang.
\newblock {Contrast enhancement via shaped Raman pulses for thermal cold atom
  cloud interferometry}.
\newblock {\em European Physical Journal D}, 70(12):262, dec 2016.

\bibitem{Fang2018}
Bess Fang, Nicolas Mielec, Denis Savoie, Matteo Altorio, Arnaud Landragin, and
  Remi Geiger.
\newblock {Improving the phase response of an atom interferometer by means of
  temporal pulse shaping}.
\newblock {\em New Journal of Physics}, 20(2):023020, feb 2018.

\bibitem{Baum1985}
J.~Baum, R.~Tycko, and A.~Pines.
\newblock {Broadband and adiabatic inversion of a two-level system by
  phase-modulated pulses}.
\newblock {\em Physical Review A}, 32(6):3435--3447, dec 1985.

\bibitem{Kovachy2012}
Tim Kovachy, Sheng~Wey Chiow, and Mark~A. Kasevich.
\newblock {Adiabatic-rapid-passage multiphoton Bragg atom optics}.
\newblock {\em Physical Review A - Atomic, Molecular, and Optical Physics},
  86(1):011606, jul 2012.

\bibitem{Bateman2007}
James Bateman and Tim Freegarde.
\newblock {Fractional adiabatic passage in two-level systems: Mirrors and beam
  splitters for atomic interferometry}.
\newblock {\em Physical Review A - Atomic, Molecular, and Optical Physics},
  76(1):013416, jul 2007.

\bibitem{Levitt1981}
Malcolm~H. Levitt and Ray Freeman.
\newblock {Composite pulse decoupling}.
\newblock {\em Journal of Magnetic Resonance (1969)}, 43(3):502--507, jun 1981.

\bibitem{Levitt1983}
Malcolm~H. Levitt and R.~R. Ernst.
\newblock {Composite pulses constructed by a recursive expansion procedure}.
\newblock {\em Journal of Magnetic Resonance (1969)}, 55(2):247--254, nov 1983.

\bibitem{Cummins2000}
H.~K. Cummins and J.~A. Jones.
\newblock {Use of composite rotations to correct systematic errors in NMR
  quantum computation}.
\newblock {\em New Journal of Physics}, 2:6--6, mar 2000.

\bibitem{Berg2015}
P.~Berg, S.~Abend, G.~Tackmann, C.~Schubert, E.~Giese, W.~P. Schleich, F.~A.
  Narducci, W.~Ertmer, and E.~M. Rasel.
\newblock {Composite-Light-Pulse Technique for High-Precision Atom
  Interferometry}.
\newblock {\em Physical Review Letters}, 114(6):063002, feb 2015.

\bibitem{Shaka1983}
A.~J. Shaka, James Keeler, Tom Frenkiel, and Ray Freeman.
\newblock {An improved sequence for broadband decoupling: WALTZ-16}.
\newblock {\em Journal of Magnetic Resonance (1969)}, 52(2):335--338, apr 1983.

\bibitem{Khaneja2005}
Navin Khaneja, Timo Reiss, Cindie Kehlet, Thomas Schulte-Herbr{\"{u}}ggen, and
  Steffen~J. Glaser.
\newblock {Optimal control of coupled spin dynamics: Design of NMR pulse
  sequences by gradient ascent algorithms}.
\newblock {\em Journal of Magnetic Resonance}, 172(2):296--305, feb 2005.

\bibitem{Hogben2011}
H.~J. Hogben, M.~Krzystyniak, G.~T~P Charnock, P.~J. Hore, and Ilya Kuprov.
\newblock {Spinach - A software library for simulation of spin dynamics in
  large spin systems}.
\newblock {\em Journal of Magnetic Resonance}, 208(2):179--194, 2011.

\bibitem{Shore2011}
B.~W. Shore.
\newblock {\em {Manipulating Quantum Structures Using Laser Pulses}}.
\newblock Cambridge University Press, 2011.

\bibitem{Tannor07}
David~J. Tannor.
\newblock {\em {Introduction to Quantum Mechanics: A Time-Dependent
  Perspective}}.
\newblock University Science Books, 2007.

\bibitem{Stoner2011}
Richard Stoner, David Butts, Joseph Kinast, and Brian Timmons.
\newblock {Analytical framework for dynamic light pulse atom interferometry at
  short interrogation times}.
\newblock {\em Journal of the Optical Society of America B}, 28(10):2418, oct
  2011.

\bibitem{Skinner2003}
Thomas~E. Skinner, Timo~O. Reiss, Burkhard Luy, Navin Khaneja, and Steffen~J.
  Glaser.
\newblock {Application of optimal control theory to the design of broadband
  excitation pulses for high-resolution NMR}.
\newblock {\em Journal of Magnetic Resonance}, 163(1):8--15, jul 2003.

\bibitem{VanFrank2014}
S.~van Frank, A.~Negretti, T.~Berrada, R.~B{\"{u}}cker, S.~Montangero, J.-F.
  Schaff, T.~Schumm, T.~Calarco, and J.~Schmiedmayer.
\newblock {Interferometry with non-classical motional states of a
  Bose–Einstein condensate}.
\newblock {\em Nature Communications}, 5, may 2014.

\bibitem{Jager2014}
Georg J{\"{a}}ger, Daniel~M. Reich, Michael~H. Goerz, Christiane~P. Koch, and
  Ulrich Hohenester.
\newblock {Optimal quantum control of Bose-Einstein condensates in magnetic
  microtraps: Comparison of gradient-ascent-pulse-engineering and Krotov
  optimization schemes}.
\newblock {\em Physical Review A - Atomic, Molecular, and Optical Physics},
  90(3):033628, sep 2014.

\bibitem{Koch2004}
Christiane~P. Koch, Jos{\'{e}}~P. Palao, Ronnie Kosloff, and Fran{\c{c}}oise
  Masnou-Seeuws.
\newblock {Stabilization of ultracold molecules using optimal control theory}.
\newblock {\em Physical Review A}, 70(1):013402, jul 2004.

\bibitem{Nobauer2015}
Tobias N{\"{o}}bauer, Andreas Angerer, Bj{\"{o}}rn Bartels, Michael Trupke,
  Stefan Rotter, J{\"{o}}rg Schmiedmayer, Florian Mintert, and Johannes Majer.
\newblock {Smooth Optimal Quantum Control for Robust Solid-State Spin
  Magnetometry}.
\newblock {\em Physical Review Letters}, 115(19):190801, nov 2015.

\bibitem{Floether2012}
Frederik~F. Floether, Pierre {De Fouquieres}, and Sophie~G. Schirmer.
\newblock {Robust quantum gates for open systems via optimal control: Markovian
  versus non-Markovian dynamics}.
\newblock {\em New Journal of Physics}, 14(7):073023, jul 2012.

\bibitem{Goodwin2015a}
D.~L. Goodwin and Ilya Kuprov.
\newblock {Auxiliary matrix formalism for interaction representation
  transformations, optimal control, and spin relaxation theories}.
\newblock {\em Journal of Chemical Physics}, 143(8):084113, aug 2015.

\bibitem{Goodwin2015}
D.~L. Goodwin and Ilya Kuprov.
\newblock {Modified Newton-Raphson GRAPE methods for optimal control of spin
  systems}.
\newblock {\em The Journal of Chemical Physics}, 144(20):204107, may 2016.

\bibitem{DeFouquieres2011}
P.~{De Fouquieres}, S.~G. Schirmer, S.~J. Glaser, and Ilya Kuprov.
\newblock {Second order gradient ascent pulse engineering}.
\newblock {\em Journal of Magnetic Resonance}, 212(2):412--417, oct 2011.

\bibitem{Levitt2008}
Malcolm~H. Levitt.
\newblock {\em {Spin Dynamics: Basic Principles of Nuclear Magnetic
  Resonance}}.
\newblock John Wiley and Sons Ltd, 2nd edition, 2008.

\bibitem{Kobzar2004}
Kyryl Kobzar, Thomas~E. Skinner, Navin Khaneja, Steffen~J. Glaser, and Burkhard
  Luy.
\newblock {Exploring the limits of broadband excitation and inversion pulses}.
\newblock {\em Journal of Magnetic Resonance}, 170(2):236--243, oct 2004.

\bibitem{Kobzar2012}
Kyryl Kobzar, Sebastian Ehni, Thomas~E. Skinner, Steffen~J. Glaser, and
  Burkhard Luy.
\newblock {Exploring the limits of broadband 90° and 180° universal rotation
  pulses}.
\newblock {\em Journal of Magnetic Resonance}, 225:142--160, dec 2012.

\bibitem{Carey2017}
Max Carey, Mohammad Belal, Matthew Himsworth, James Bateman, and Tim Freegarde.
\newblock {Matterwave interferometric velocimetry of cold Rb atoms}.
\newblock {\em Journal of Modern Optics}, 65(4):1--10, feb 2017.

\bibitem{Dunning2014a}
Alex Dunning.
\newblock {\em {Coherent atomic manipulation and cooling using composite
  optical pulse sequences}}.
\newblock PhD thesis, University of Southampton, UK, 2014.

\bibitem{Kobzar2008a}
Kyryl Kobzar, Thomas~E. Skinner, Navin Khaneja, Steffen~J. Glaser, and Burkhard
  Luy.
\newblock {Exploring the limits of broadband excitation and inversion: II.
  Rf-power optimized pulses}.
\newblock {\em Journal of Magnetic Resonance}, 194(1):58--66, 2008.

\bibitem{Ryan2010}
C.~A. Ryan, J.~S. Hodges, and D.~G. Cory.
\newblock {Robust Decoupling Techniques to Extend Quantum Coherence in
  Diamond}.
\newblock {\em Physical Review Letters}, 105(20):200402, nov 2010.

\bibitem{Luo2016a}
Yukun Luo, Shuhua Yan, Jun Yang, Qingqing Hu, Aiai Jia, Chunhua Wei, and
  Guochao Wang.
\newblock {Global performance investigation of composite pulses in atom
  interferometry}.
\newblock {\em 2015 11th Conference on Lasers and Electro-Optics Pacific Rim,
  CLEO-PR 2015}, 2(2):2--3, 2016.

\bibitem{Braun2010}
Michael Braun and Steffen~J. Glaser.
\newblock {Cooperative pulses}.
\newblock {\em Journal of Magnetic Resonance}, 207(1):114--123, nov 2010.

\bibitem{Braun2014}
Michael Braun and Steffen~J. Glaser.
\newblock {Concurrently optimized cooperative pulses in robust quantum control:
  application to broadband Ramsey-type pulse sequence elements}.
\newblock {\em New Journal of Physics}, 16(11):115002, oct 2014.

\bibitem{Levitt1981a}
Malcolm~H Levitt and Ray Freeman.
\newblock {Compensation for pulse imperfections in NMR spin-echo experiments}.
\newblock {\em Journal of Magnetic Resonance (1969)}, 43(1):65--80, apr 1981.

\bibitem{Cummins2003}
Holly~K. Cummins, Gavin Llewellyn, and Jonathan~A. Jones.
\newblock {Tackling systematic errors in quantum logic gates with composite
  rotations}.
\newblock {\em Physical Review A}, 67(4):042308, apr 2003.

\bibitem{Wimperis1994}
S.~Wimperis.
\newblock {Broadband, Narrowband, and Passband Composite Pulses for Use in
  Advanced NMR Experiments}.
\newblock {\em Journal of Magnetic Resonance, Series A}, 109(2):221--231, aug
  1994.

\end{thebibliography}

\end{document}